\documentclass[reprint,floatfix]{revtex4-2}
\usepackage[utf8]{inputenc}
\usepackage[T1]{fontenc}
\usepackage[english]{babel}
\usepackage{import}
\usepackage{amsmath}
\usepackage{amsfonts}
\usepackage{amssymb}
\usepackage{bm}
\usepackage{graphicx}
\usepackage{braket}
\usepackage{hyperref}
\usepackage{natbib}
\usepackage[table,xcdraw]{xcolor}

\def\*#1{\mathbf{#1}}

\begin{document}
\preprint{APS/123-QED}

\title{Mapping of a many-qubit state onto an oscillator using controlled displacements}

\author{Anders J. E. Bjerrum}
    \affiliation{Center for Macroscopic Quantum States (bigQ), Department of Physics, Technical University of Denmark, 2800 Kongens Lyngby, Denmark}
\author{Ulrik L. Andersen}%
    \affiliation{Center for Macroscopic Quantum States (bigQ), Department of Physics, Technical University of Denmark, 2800 Kongens Lyngby, Denmark}
\author{Peter Rabl}%
    \affiliation{$^2$Walther-Mei\ss ner-Institut, Bayerische Akademie der Wissenschaften, 85748 Garching, Germany}

\affiliation{$^3$Technische Universit\"at M\"unchen, TUM School of Natural Sciences, Physics Department, 85748 Garching, Germany} 

\affiliation{$^4$Munich Center for Quantum Science and Technology (MCQST), 80799 Munich, Germany}
\date{\today}

\begin{abstract}
We extend the controlled displacement interaction between a qubit and a harmonic oscillator to the multi-qubit (qudit) case. We define discrete quadratures of the qudit and show how the qudit state can be displaced in these quadratures controlled by an oscillator quadrature. Using this interaction, a periodic repetition of the state encoded in the qudit, can be deterministically mapped onto the oscillator, which is initialized in a squeezed state.
Based on this controlled displacement interaction, we present a full circuit that encodes quantum information in a superposition of qudit quadrature states, and successively prepares the oscillator in the corresponding superposition of approximate Gottesman-Kitaev-Preskill (GKP) states. This preparation scheme is found to be similar to phase estimation, with the addition of a disentanglement gate.
Our protocol for GKP state preparation is efficient in the sense, that the set of qubits need only interact with the oscillator through two time-independent interactions, and in the sense that the squeeze factor (in dB) of the produced GKP state grows linearly in the number of qubits used.
\end{abstract}
\maketitle
\section{Introduction}
\label{chapter:gkp:introTitle}
We present a protocol for deterministically mapping a state encoded in a set of qubits onto an oscillator. This mapping is utilized to generate approximate Gottesman-Kitaev-Preskill (GKP) states \cite{Gottesman:2001}, which have been given much attention in the literature as a promising candidate for encoding an error correctable qubit in an oscillator \cite{Grimsmo:2021,Walshe:2020}. These states can be protected against small shifts in phase space, and will be relevant to implementations of bosonic quantum computing \cite{Bourassa:2021,Baragiola:2019} and also quantum communication \citep{Jiang:2016,Fukui:2021,Guha:2024:1,Guha:2024:2}.\\
Various methods for preparing GKP states exist in the literature including Floquet engineering \cite{Kolesnikow:2024}, cavity QED \cite{Hastrup:2021,Hastrup2:2022}, cat breeding \cite{Konno:2024,Weigand:2018}, boson sampling \cite{Tzitrin:2020,Takase:2023}, phase estimation \cite{Terhal:2016,Shi:2019}, the dispersive Faraday interaction \cite{Motes:2017}, and shaped free electrons \cite{Dahan:2023}. We note that experimental realizations of approximate GKP states have been achieved in trapped ion systems \cite{Fluhmann:2019,Neeve:2022}, and superconducting microwave cavities \cite{Campagne:2020,Eickbusch:2022}. In this work we present a protocol for preparing GKP states which utilize a many-qubit state. By leveraging the rich structure of multiple qubits, we show that by tailoring the qudit state we can efficiently prepare approximate GKP states of the oscillator. Efficient in the sense that the squeeze factor (in dB) of the GKP state grows linearly in the number of employed qubits, and in the sense that the set of qubits need only interact with the oscillator through two time-independent interactions. Our scheme is highly related to phase estimation, however our protocol does not require any measurements to be performed. We expect that the scheme proposed in this work could be implemented using superconducting microwave cavities as in \cite{Eickbusch:2022}, but also in other platforms.


The non-commutativity between the $q$ and $p$ quadrature operators, as dictated by the canonical commutation relation of quantum optics $[q,p] = i$ (setting the vacuum variance to 1/2) enforces that no common eigenstates exist for $q$ and $p$. However, the corresponding modular observables $q_m = q \pmod{Q}$ and $p_m = p \pmod{\frac{2\pi}{Q}}$ posses common eigenstates \citep{Gottesman:2001,Aharonov:2008} for any real number $Q$, and these states can be written as,
\begin{align}
	\ket{q_0,p_0} = \sum_{s=-\infty}^{\infty}e^{ip_0(q_0-sQ)}|q_0-sQ\rangle ,
\end{align}
with $q_0$ and $p_0$ being the eigenvalues of $q_m$ and $p_m$ respectively. We note that $\ket{q_0,p_0}$ is not normalizable, and does not constitute a proper quantum state. 
However, these states can be closely approximated, with GKP states providing a practical representation. The GKP states can be expressed as,
\begin{align}
	\ket{G(\phi)}&=\mathcal{N}_0\sum_{s=-\infty}^{\infty}e^{-(\kappa(2s+\phi)\sqrt{\pi})^{2}/2} \nonumber \\ &\int_{-\infty}^{\infty} dq \pi^{-1/4}\Delta^{-1/2} e^{-q^{2}/(2\Delta^{2})} \ket{q+(2s+\phi)\sqrt{\pi}},
	\label{intro:eq:approxGKP}
\end{align}
where $\mathcal{N}_0$ is the normalization. In position space, these states represent a train of peaks of width $\Delta$ and spacing $2\sqrt{\pi}$. The parameter $\kappa$ determines the overall gaussian envelop, and $\phi$ shifts the center of the GKP state.  By choosing $\phi= 0$ and $\phi =1$ we obtain two orthogonal states (for small enough $\Delta$), which can be identified with the logical qubit states $|0_L\rangle\equiv |G(0)\rangle$ and $|1_L\rangle\equiv |G(1)\rangle$, respectively. \\

Starting from of an oscillator prepared in a squeezed vacuum state, the central task is to prepare the oscillator in a superposition of displaced gaussian wave packets, which represents an arbitrary qubit state $|\psi_L\rangle= \alpha |0_L\rangle +\beta |1_L\rangle$. Previous works have demonstrated \citep{Travaglione:2002,Terhal:2016,Hastrup:2021,Eickbusch:2022} that the preparation of GKP states can be achieved using controlled displacement interactions between a qubit and an oscillator. Controlled displacements allow us to displace the oscillator in the $p$ quadrature, conditional on the spin state of the qubit, and are generated by the Hamiltonian,
\begin{align}
	H_{\mathrm{int}}/\hbar = \chi q \sigma_z
\end{align}
where $\sigma_z$ is the Pauli z operator for the qubit, and $\chi$ is a coupling frequency. We have the $\sigma_z$ eigenstates $\sigma_z|0\rangle = |0\rangle$ and $\sigma_z|1\rangle = -|1\rangle$. This approach requires a repeated interaction between qubit and oscillator, and after each step the qubit must be disentangled from the resonator either via a projective measurement or a disentangling gate. Imperfections in these operations as well as the decoherence of the qubit during the interaction limit the achievable fidelity of the prepared state \cite{Eickbusch:2022}.

In the following we present a different preparation strategy where the sequential interaction with a single qubit is replaced by a single parallel interaction with $N>1$ qubits. To do so we consider a qubit-oscillator coupling of the form 
\begin{equation} 
H_{\rm int}/\hbar = \chi q X_N,
\end{equation} 
where $X_N = -\sum_{n=1}^N 2^{n-2} \sigma_z^{(n)}$ is the discrete quadrature operator in the $N$-qubit subspace. We are not aware of any existing experimental demonstration of such an interaction. A difficulty associated with this interaction is that multiple qubits would have to couple to the same oscillator with varying coupling rates. By preparing an appropriate initial state of the qubits, this interaction allows us to prepare an arbitrary encoded state $|\psi_L\rangle$ by a single global controlled displacement gate and one global disentangling operation. In this way, the interaction time, during which the state is sensitive to qubit dephasing, is minimized and most of the gates for preparing the required initial qubit state can be implemented while the oscillator and the qubits are decoupled. \\

Given an initial oscillator state with $q$-quadrature wavefunction $\psi(q) = \langle q | \psi \rangle \propto \exp\left(-\frac{q^{2}}{2W^{2}}\right)$ of width $W>1$ (anti-squeezed in $q$), our protocol will produce GKP states with an envelope,
\begin{equation}
\kappa = \frac{1}{W},
\end{equation} 
while the width of the individual peaks decreases exponentially with the number of qubits, 
\begin{equation}
 \Delta \propto  2^{-N},
\end{equation} 
with a proportionality constant of around $3.1$. For $N=3$ and $N=4$ qubits, we find that the protocol ideally produces approximate GKP states with $\Delta \approx 0.39$ (8.2 dB) and $\Delta \approx 0.19$ (14.4 dB), respectively. 


\subsection{Qudit quadratures}
Before we describe the actual protocol, we start by outlining the underlying theory. We define discrete quadrature operators for a qudit \citep{Nielsen:2010,Santhanam:1976}. We then realize the qudit using a set of qubits, and the qudit quadratures are then constructed from the Hamiltonian of this system. Finally, we present a unitary that displaces the qudit in one of the qudit quadratures, conditioned on an oscillator quadrature. \\
Given a qudit of dimension $M$, we denote the qudit levels by the basis states $\ket{x_k}$ where $k$ is a number belonging to the set,
\begin{align}
	k \in K =  \Biggl \{ \pm1/2,\pm3/2,\cdots, \pm(M-1)/2 \Biggl \} .
\end{align}
The basis states $\ket{x_k}$ can for example be realized as the logical basis states of a set of qubits, as we will show in the next section. Using these states we introduce a discrete operator,
\begin{align}
	X_N = \sum_{k \in K} k \ket{x_k} \bra{x_k} .
\end{align}
We refer to $X_N$ as a quadrature operator because it has a linear spectrum with both positive and negative values. Applying the discrete Fourier transform, i.e. the quantum Fourier transform $F_N$, we can transform $X_N$ into a conjugate quadrature operator,
\begin{align}
	Y_N = F_N X_N F_N^\dagger
\end{align}
where
\begin{align}
	F_N = \frac{1}{\sqrt{M}} \sum_{n,m \in K} e^{i2\pi (n-1/2) (m-1/2) /M} \ket{x_m}\bra{x_n} ,
	\label{protocol:eq:qft}
\end{align}
and $Y_N$ has eigenvectors,
\begin{align}
	\ket{y_n} = \frac{1}{\sqrt{M}} \sum_{k \in K} e^{i2\pi (n-1/2)(k-1/2)/M} \ket{x_k} ,
\end{align}
with eigenvalues $n \in K$. \\
We will now argue that $Y_N$ is the generator of translations in the qudit levels. To see this we introduce the qudit state $\ket{v}$ with amplitudes $v_k$,
\begin{align}
	\ket{v} = \sum_{k \in K} v_k \ket{x_k},
\end{align}
and note that the following expression interpolates the points $(k,v_k)$ over the interval $y\in R_0 = [-M/2,M/2[$,
\begin{align}
	v(y)=\frac{1}{M}\sum_{n,l\in K}v_{l}e^{i2\pi (n-1/2)\left(y-l\right)/M} ,
	\label{modmeas:equation:interpolation}
\end{align}
as can be verified by evaluating $v \hspace{-0.1cm} \left( k \right)$ for $k \in K$. Extending the range of $y$ outside $R_0$ shows that $v(y)$ repeats itself with period $M$. More details on $v(y)$ can be found in appendix \ref{appendix:section:interpolater}. \\
We then exponentiate $Y_N$ as $D_x(s) = e^{-i \frac{2\pi}{M} Y_N s}$ where $s$ is an arbitrary real number. Evaluating the action of $D_x(s)$ on the the state $\ket{v}$, we obtain,
\begin{align}
	D_{x}(s)\sum_{k\in K}v_{k}|x_{k}\rangle = e^{-i\frac{\pi}{M}s}\sum_{m\in K} v(m-s) |x_{m}\rangle .
	\label{protocol:eq:displacement}
\end{align}
Hence the action of $D_{x}(s)$ on $|v\rangle$ is to generate a new qudit state, where the amplitudes of this new state are obtained by sampling from a translation of the interpolating function $v(y)$. 

\subsubsection{Qubit realization}
\begin{figure*}
	\centering		
	\includegraphics[width=0.9\textwidth]{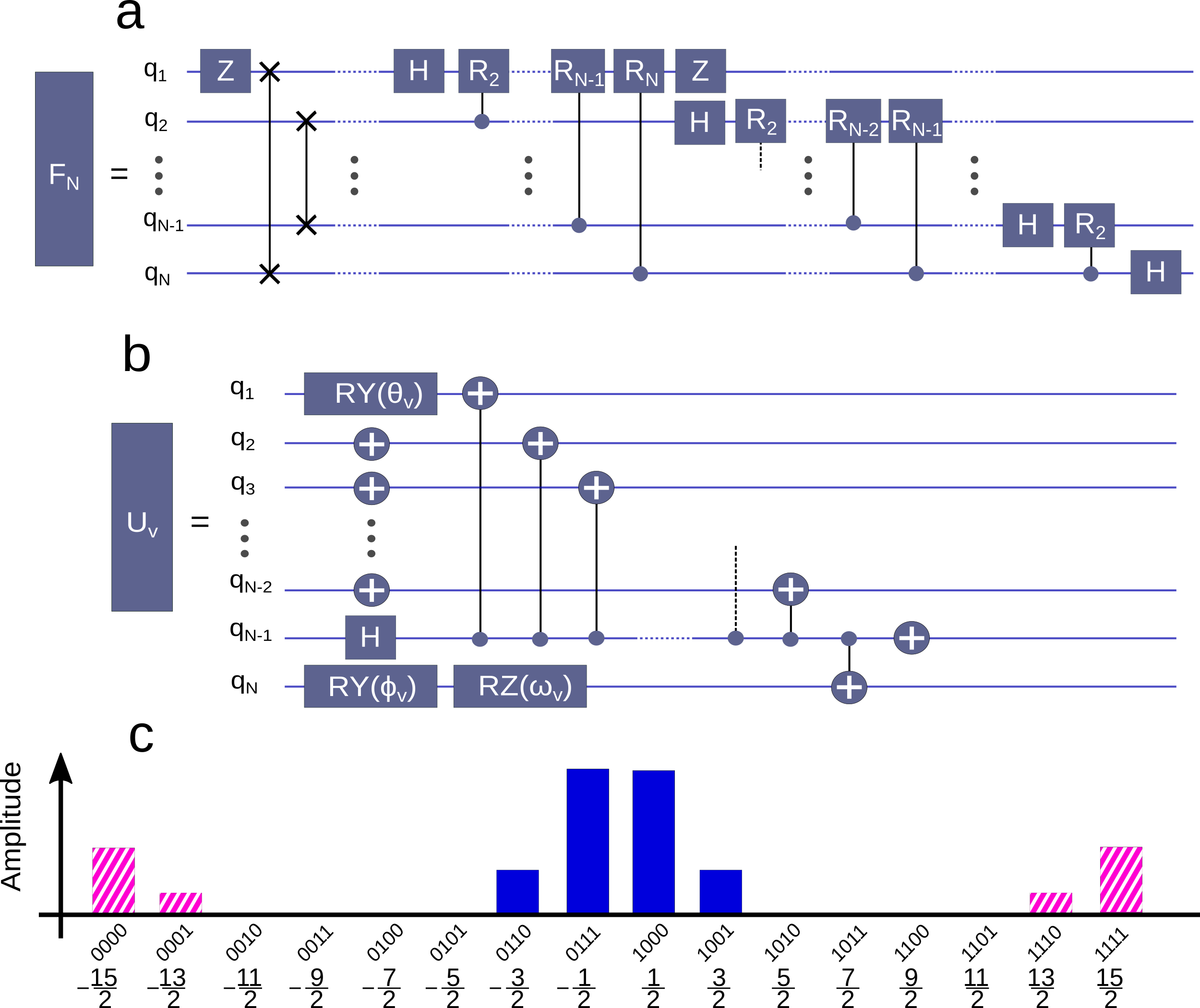}
		\caption{\textit{\textbf{a} Quantum circuit for the version of the quantum Fourier transform used in this work. The connected crosses indicate a swap gate. \textbf{b} Circuit for preparing the family of qubit states $\ket{v}$. The circuit corresponds to the unitary $U_v$. The encircled plus symbol is a Pauli x gate. The controlled gates are controlled not gates, and the target qubit is denoted by an encircled plus.  $\theta_v$ is chosen so that the resulting state resembles a gaussian distribution, we find that $\theta_v = 2.6$ works well. This distribution can be brought into an arbitrary superposition of two central positions by tuning the relative amplitude via $\phi_v$ and the relative phase via $\omega_v$. \textbf{c} A sketch of a quantum state generated by the unitary $U_v$ given in b, with 4 qubits initially in the state $\ket{0,0,0,0}$, and with $\theta_v = 2.6$. On the x-axis we give the logical state of the qubits, and the associated eigenvalue of the $X_N$ operator. On the y-axis we sketch the amplitudes, with color indicating phase. If we impose periodic boundary conditions, we note that the state is a superposition of two localized peaks. The relative amplitude and phase of these peaks are controlled by $\phi_v$ and $\omega_v$ respectively. }}
		\label{modsmeas:fig:UFCircuit}
\end{figure*}

Given a collection of $N$ qubits we form a qudit of dimension $M=2^{N}$. The qudit levels are now the associated logical basis states, $\ket{x_k} = \ket{j_{N},j_{N-1},\hdots,j_2,j_1}$, where $j_n\in\{0,1\}$. The label $k$ is related to the logical state as,
\begin{align}
	k =	\sum_{n=1}^{N} j_n 2^{n-1} - (M-1)/2 .
\end{align}
The circuit for generating the QFT $F_N$ is shown in Fig. \ref{modsmeas:fig:UFCircuit} a \cite{Nielsen:2010}. This circuit uses the gate $R_k$ which can be written in the $\sigma_z$-basis $\left( \ket{0}=\begin{pmatrix} 1 & 0\end{pmatrix}^T \right)$ as
\begin{align}
	R_k = \begin{pmatrix}
		1 & 0 \\
		0 & \mathrm{e}^{2\pi i/2^k}
	\end{pmatrix}.
\end{align}
We construct the quadrature operator $X_N$ using the $\sigma_z^{(n)}$ operators,
\begin{align}
	X_N = -\sum_{n=1}^{N} 2^{n-2} \sigma_z^{(n)}
	\label{prtcl2:eq:xnQ}
\end{align}
where $\sigma_z^{(n)}$ is applied to the $n$-th qubit. Assuming we have access to a controlled displacement between each of the qubits and an oscillator, we assume a qubit-oscillator coupling of the form 
\begin{equation} 
H_{\rm int}/\hbar = \chi q X_N = - \chi \sum_{n=1}^{N} 2^{n-2} q \sigma_z^{(n)},
\label{prtcl2:eq:qX}
\end{equation} 
where $q$ is a quadrature of the oscillator. Applying the QFT and evolving under the time-independent Hamiltonian $H_{\rm int}/\hbar$ for a time $\tau_I$, we observe the following result,
\begin{align}
	F_N e^{i \chi \tau_I q X_N } F_N^\dagger = e^{i \chi \tau_I q Y_N } = D_x \left(- \frac{M  q}{P_q} \right),
\end{align}
where we have defined the dimensionless parameter $P_q = 2\pi /(\chi \tau_I)$.
Hence if we have access to controlled displacements and the QFT, then we can displace the qubit state through the $X_N$ quadrature, conditioned on the quadrature $q$ of an oscillator. \\

\begin{figure*}
	\centering		
	\includegraphics[width=1\textwidth]{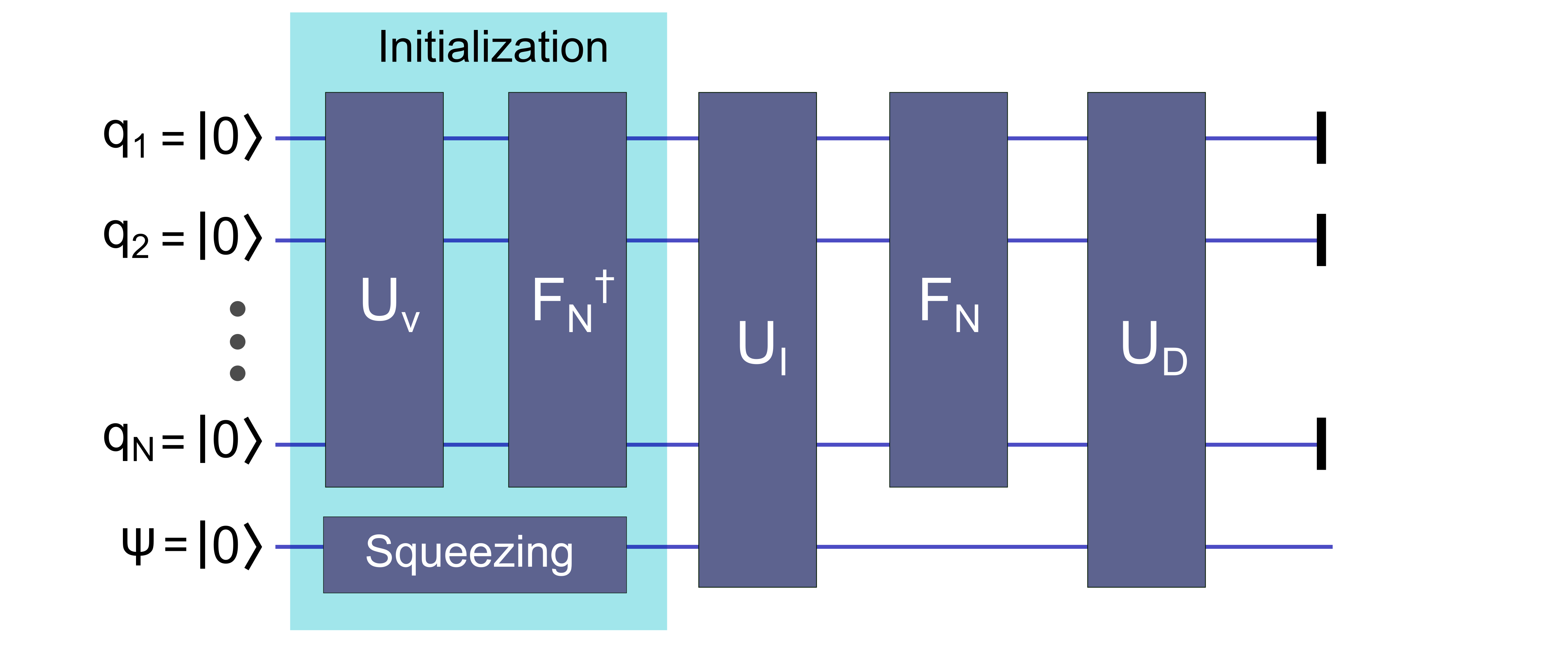}
		\caption{\textit{Sketch of the protocol given in the main text. The circuit $U_v$ generates the initial qubit state. We apply the inverse QFT on the qubits, and we then apply the unitary $U_I$. The QFT is then applied on the qubits, and a disentangling operation $U_D$ follows.}}
		\label{modsmeas:fig:Prtcl2Circuit}
\end{figure*}
\section{The protocol}
We can now describe our protocol which is sketched in Fig. \ref{modsmeas:fig:Prtcl2Circuit}. The protocol proceeds as follows,
\begin{enumerate}
	\item{The initial oscillator state $\psi_0(q)$ is a displaced squeezed state in the $q$ quadrature, $\psi_0(q) = e^{-iq\pi/P_q}\psi_W(q)$, where $\psi_W(q) = C \exp\left(-\frac{q^{2}}{2W^{2}}\right)$ is squeezed vacuum of width $W>1$, i.e. we have anti-squeezing.}
	\item{The qubits are initialized in the state $|v\rangle = U_v|0,0,0,...,0\rangle$.}
	\item{The inverse QFT is applied to the qubits}
	\item{The qubits interact with the oscillator through the unitary $U_I = e^{i  (2\pi/P_q) q X_N}$.}
	\item{The QFT is applied to the qubits}
	\item{The qubits interact with the oscillator through $U_D = e^{- i (P_q/M) p X_{N}}$. }
\end{enumerate}
To implement our protocol we prepare the qubit state $\ket{v}$ using the circuit shown in Fig. \ref{modsmeas:fig:UFCircuit} b which acts on the input state $\ket{0,0,\hdots,0}$. A sketch of the amplitudes associated with the state $\ket{v}$ is shown in Fig. \ref{modsmeas:fig:UFCircuit} c. The final result of our protocol is to map the interpolation $v(y)$ of the state $\ket{v}$ onto the oscillator in a periodic fashion. The state $\ket{v}$ is therefore chosen such that its periodic continuation yields a train of peaks. In particular we aim for $v(y)$ to produce two sets of peaks: One set that is associated with the logical 0 GKP state, and one set that is associated with the logical 1 GKP state. The circuit shown in Fig. \ref{modsmeas:fig:UFCircuit} b is denoted by the unitary $U_v$, so we may write $|v\rangle = U_v |0,0,\hdots,0\rangle$. We have defined the operators $\mathrm{RY}(\phi_v)=e^{-i(\phi_v/2)\sigma_{y}}$ and $\mathrm{RZ}(\omega_v)=e^{-i(\omega_v/2)\sigma_{z}}$. The angles $\phi_v$ and $\omega_v$ are used to create an arbitrary superposition of two central positions, as shown in Fig. \ref{modsmeas:fig:UFCircuit} c, with $\phi_v$ controlling the relative amplitude and $\omega_v$ controlling the relative phase, such that the protocol prepares the GKP state  $|\Psi_L\rangle = \cos(\phi_v/2) |0_L\rangle +\sin(\phi_v/2) e^{i\omega_v} |1_L\rangle$. We set $\theta_v = 2.6$ as this generates a qubit state with approximately gaussian peaks, as sketched in Fig. \ref{modsmeas:fig:UFCircuit} c. For $\theta_v$ outside the range $[2.5,2.7]$ the function $v(y)$, which interpolates the qubit state, will have a significant amount of undesirable oscillations.
Breaking down the protocol, we examine the following term,
\begin{align}
	F_N U_I F_N^\dagger \ket{v} \ket{\psi_0} &= \int_{-\infty}^{\infty} dq  D_x \hspace{-0.1cm} \left( -\frac{M }{P_q} q \right)   \ket{v} \psi_0(q) \ket{q} \nonumber \\
	&= \sum_{l\in K} \int_{-\infty}^{\infty} dq  v\hspace{-0.1cm} \left( l + \frac{M }{P_q} q \right) \psi_W(q) \ket{x_l} \ket{q} ,
	\label{prtcl2:eq:entangled}
\end{align}
showing that the interaction has the effect of modulating the squeezed vacuum $\psi_W(q)$ by the periodic function $ v\hspace{-0.1cm}\left( l+\frac{M }{P_q}  q \right)$ with period $P_q $ in $q$. \\

\begin{figure*}
	\centering		
	\includegraphics[width=1\textwidth]{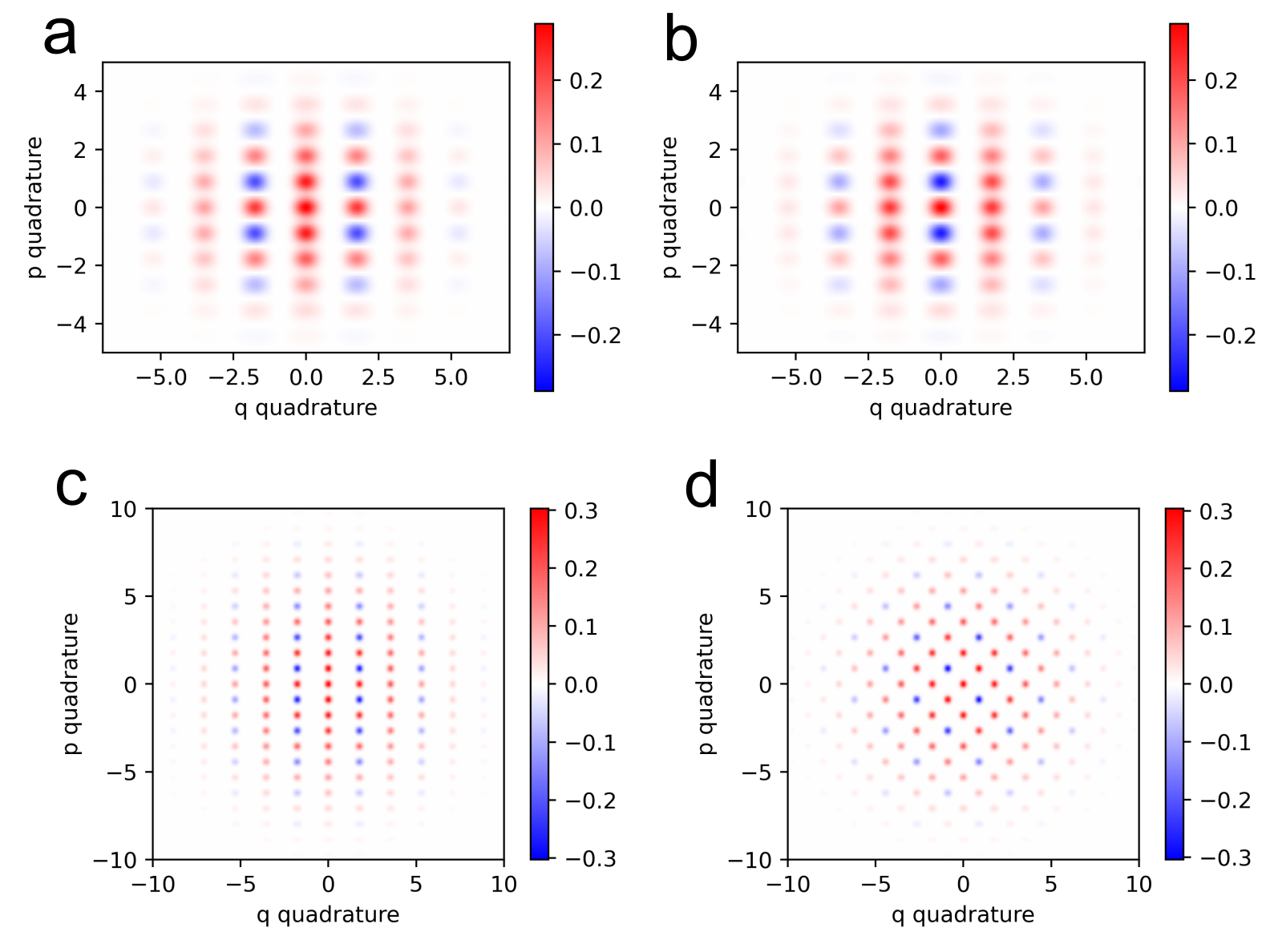}
		\caption{\textit{Wigner functions for states created by the protocol. \textbf{a} Wigner function for 3 qubits with 10 dB of initial squeezing, i.e. $W=3.2$. Furthermore we have $\phi_v=\omega_v=0$ and $P_q = 2\sqrt{\pi}$. The Wigner function is seen to strongly resemble a logical 0 GKP state. \textbf{b} Wigner function for 3 qubits with 10 dB of initial squeezing, $\phi_v=\pi$ and $\omega_v=0$. The Wigner function is seen to strongly resemble a logical 1 GKP state. \textbf{c} Wigner function for 4 qubits with 14 dB of initial squeezing and $\phi_v=\omega_v=0$. \textbf{d} Wigner function for 4 qubits with 14 dB of initial squeezing and $\phi_v = \pi/2$ and $\omega_v = \pi/2$.}}
		\label{modsmeas:fig:wigner}
\end{figure*}

Next, we apply the unitary $U_D$ on the qubits and the oscillator, and we will see that $U_D$ approximately disentangles the qubits and oscillator. Applying $U_D$ we get,
\begin{align}
	&U_D F_N U_I F_N^\dagger \ket{v} \ket{\psi_0} \nonumber \\
	&= \int_{-\infty}^{\infty} dq \psi_W(q) \sum_{l\in K} v\hspace{-0.1cm}\left( l+\frac{M}{P_q} q \right) e^{-i\frac{P_q}{M} p X_{N} } \ket{x_l} \ket{q} \nonumber \\
	&= \int_{-\infty}^{\infty} dq \psi_W(q) \sum_{l\in K}  v\hspace{-0.1cm}\left( l+\frac{M}{P_q} q \right) \ket{x_l} \biggl | q + l \frac{P_q}{M} \biggl \rangle \nonumber \\
	&= \int_{-\infty}^{\infty} dq \sum_{l\in K} \psi_W\hspace{-0.1cm}\left( q - l \frac{P_q}{M}  \right)   v\hspace{-0.1cm}\left( \frac{M}{P_q} q \right) \ket{x_l} \ket{q} 	.	
	\label{prtcl2:eq:disentanglement}
\end{align}
Provided that $\psi_W(q)$ varies slowly in $q$, so that it does not change signficantly over the period $P_q$, then we observe that there is only a small amount of entanglement between the oscillator and the qubits in the above expression, since $l$ is bounded by $\pm M/2$. This condition holds approximately if $W > P_q/2$. Under this assumption we can rewrite Eq. \ref{prtcl2:eq:disentanglement} as,
\begin{align}
	&U_D F_N U_I F_N^\dagger \ket{v} \ket{\psi_0} \nonumber \\ &\approx \int_{-\infty}^{\infty} dq \psi_W\hspace{-0.1cm}\left( q\right)   v\hspace{-0.1cm}\left(\frac{M}{P_q} q  \right)  \ket{q} \sum_{l\in K}\ket{x_l} ,
\end{align}
showing that the oscillator state is now anti-squeezed vacuum modulated by the periodic function $v\hspace{-0.1cm}\left(\frac{M}{P_q} q  \right)$. The anti-squeezed vacuum $\psi_W\left( q\right)$ corresponds to the envelope in Eq. \ref{intro:eq:approxGKP}, whereas the width of $v\hspace{-0.1cm}\left(\frac{M}{P_q} q  \right)$ in $q$ corresponds to $\Delta$ in Eq. \ref{intro:eq:approxGKP}. The standard deviation of $v(y)$ is $\sigma = 0.83$ for $\theta_v = 2.6$. Since $M = 2^{N}$ we find that the width of $v\hspace{-0.1cm}\left(\frac{M}{P_q} q  \right)$ in $q$, and hence $\Delta$, scales as,
\begin{align}
	\Delta \approx \frac{\sigma P_q}{2^N},
\end{align}
where $\sigma P_q \approx 2.9$.
However, $v(y)$ is not perfectly gaussian, and by numerically fitting the expression in Eq. \ref{intro:eq:approxGKP} to the states generated by the protocol, we find better agreement with the relation $\Delta \approx 3.1/2^N$. It follows that the squeeze factor (in dB) of the approximate GKP state will grow linearly in the number of qubits $N$.  \\
Without invoking the above approximation we can describe the state of the oscillator using the reduced density matrix,
\begin{align}
	\rho_{\psi} = \int_{-\infty}^{\infty} dz  \int_{-\infty}^{\infty} dy & \sum_{l\in K} \psi_W \hspace{-0.1cm}\left( z - l\frac{P_q }{M} \right) \psi_W \hspace{-0.1cm}\left( y - l \frac{P_q }{M} \right)^* \nonumber \\
	&  v \hspace{-0.1cm}\left( \frac{M}{P_q} z \right) v \hspace{-0.1cm}\left( \frac{M}{P_q} y  \right)^* \ket{z}\bra{y} .
	\label{prtcl2:eq:densMatrix}
\end{align}
We plot the Wigner function of the state in Eq. \ref{prtcl2:eq:densMatrix} for various values of $N$ and $W$, using $\theta_v = 2.6$. The results are shown in Fig. \ref{modsmeas:fig:wigner}
\subsection{Implementation with a dispersive Hamiltonian}
Following the approach in \cite{Eickbusch:2022} we show how to implement the controlled displacements used in our protocol. This approach leverages a dispersive interaction between the qubits and the oscillator. In the dispersive regime, we assume the following Hamiltonian \cite{Blais:2021},
\begin{align}
	H/\hbar &=\sum_{n=1}^{N}\left\{ \chi^{(n)}a^{\dagger}a\frac{\sigma_{z}^{(n)}}{2}+\frac{1}{2}\omega_{q}^{(n)}\sigma_{z}^{(n)}\right\} \nonumber \\ &+\varepsilon^{*}(t)a+\varepsilon(t)a^{\dagger}+\omega_{o}a^{\dagger}a ,
	\label{modmeas:eq:dispHam}
\end{align}
where the drive $\varepsilon(t)$ is resonant with the oscillator. $\omega_o$ and $\omega_q^{(n)}$ are oscillator and qubit frequencies respectively. This drive will perform displacements in phase space to engineer the target controlled displacement gate between the qubits and the oscillator. We assume that the coupling frequencies can be arranged such that $\chi^{(n)} = \chi 2^{n-2}$. Considering a transmon-oscillator dispersive interaction, one finds that the coupling frequency $\chi^{(n)}$, can for example be varied by changing the detuning or charging energy of the transmon \cite{Blais:2021}. We also note the exponential scaling of the coupling frequencies $\chi^{(n)}$, a feature shared with the related standard phase estimation protocol \cite{Terhal:2016}, but which the authors circumvent by replacing standard phase estimation with phase estimation by repetition. It is crucial that the relative error on these coupling frequencies is small, ensuring that the error on the largest coupling frequency is small compared to the magnitude of the smallest coupling frequency. This required exponential scale of coupling frequencies $\chi^{(n)}$, implies that our protocol has a significant sensitivity toward the relative error on the coupling frequencies $\chi^{(n)}$.\\ 
The density matrix $\rho$ of the joint qubits-oscillator system evolves according to a master equation, given in Lindblad form as,
\begin{align}
	\partial_{t}\rho=-i [H/\hbar,\rho]+\sum_{i}\kappa_{i}D[L_{i}]\rho
\end{align}
with $D[L_i]$ being the Lindbladian associated with operator $L_i$,
\begin{align}
	D[L_i] = L_{i}\rho L_{i}^{\dagger}-\frac{1}{2}\left(L_{i}^{\dagger}L_{i}\rho+\rho L_{i}^{\dagger}L_{i}\right) ,
\end{align}
and $\kappa_i$ is the associated rate.
We include in our analysis the noise sources listed in Table \ref{modmeas:table:noise}.
\begin{center}
\begin{table}[]
\centering
\begin{tabular}{c|c|l}
Jump operator $L_i$            & Rate $\kappa_i$    & Description            \\ \hline \hline
$a$              & $\kappa_l$    & Photon loss            \\ \hline
$\sigma_-^{(n)}$ & $\gamma_l$    & Qubit decay            \\ \hline       
$a^\dagger a$    & $2\kappa_\phi$ & Oscillator phase noise \\ \hline
$\frac{1}{2} \sigma_z^{(n)}$ & $2\gamma_\phi$ & Qubit phase noise      \\
\end{tabular}
\caption{\textit{Table of noise sources and their associated jump operator and rate \citep{Agarwal:2013,Eickbusch:2022}.}}
\label{modmeas:table:noise}
\end{table}
\end{center}
Following \cite{Eickbusch:2022}, we transition to a displaced rotated frame $\mathcal{D}$, as detailed in the appendix section \ref{appendix:section:frameTrans} to \ref{appendix:section:masterEq}. In this displaced frame we  shift the origin in phase-space such that the ordinary vacuum state is a displaced coherent state. It is useful to work in a displaced frame when simulating states that are localized and centred far out in phase-space. While such states will have a high photon number in the ordinary frame, they might have a low photon number in the displaced frame. We introduce the complex number $\alpha(t)$ such that $\ket{-\alpha(t)} \bra{-\alpha(t)}$ represents the vacuum state in the ordinary frame as viewed from the displaced frame $\mathcal{D}$.
In frame $\mathcal{D}$ we have the effective Hamiltonian,
\begin{align}
H_{\mathrm{eff}}/\hbar&=\sum_{n=1}^{N}\bigg\{ \frac{1}{2}\chi^{(n)}a^{\dagger}a\sigma_{z}^{(n)} + \frac{1}{2}\chi^{(n)}|\alpha(t)|^{2}\sigma_{z}^{(n)} \nonumber \\ 
&+\frac{1}{2}\chi^{(n)}\left(\alpha(t)a^{\dagger}+\alpha^{*}(t)a\right)\sigma_{z}^{(n)} \bigg\}
\label{modmeas:eq:effH} 
\end{align}
where the drive $\varepsilon(t)$ satisfies,
\begin{align}
	e^{i\omega_o t}\varepsilon(t) - i \dot{\alpha}(t) - \frac{1}{2} i \kappa_l \alpha(t) = 0 .
\end{align}
It was demonstrated in \cite{Eickbusch:2022} that the term $\frac{1}{2}\sum_{n=1}^N \chi^{(n)} a^{\dagger}a\sigma_{z}^{(n)}$ can be suppressed by flipping the qubits and $\alpha(t)$ after half of the interaction time. By suppressing this term the effective Hamiltonian $H_{\mathrm{eff}}$ becomes the desired controlled displacement interaction. We will assume that the qubit rotation $\frac{1}{2}\chi^{(n)}|\alpha(t)|^{2}\sigma_{z}^{(n)}$ can be corrected for and it is ignored in the following analysis. 

\section{Discussion}

\begin{figure*}
	\centering		
	\includegraphics[width=0.8\textwidth]{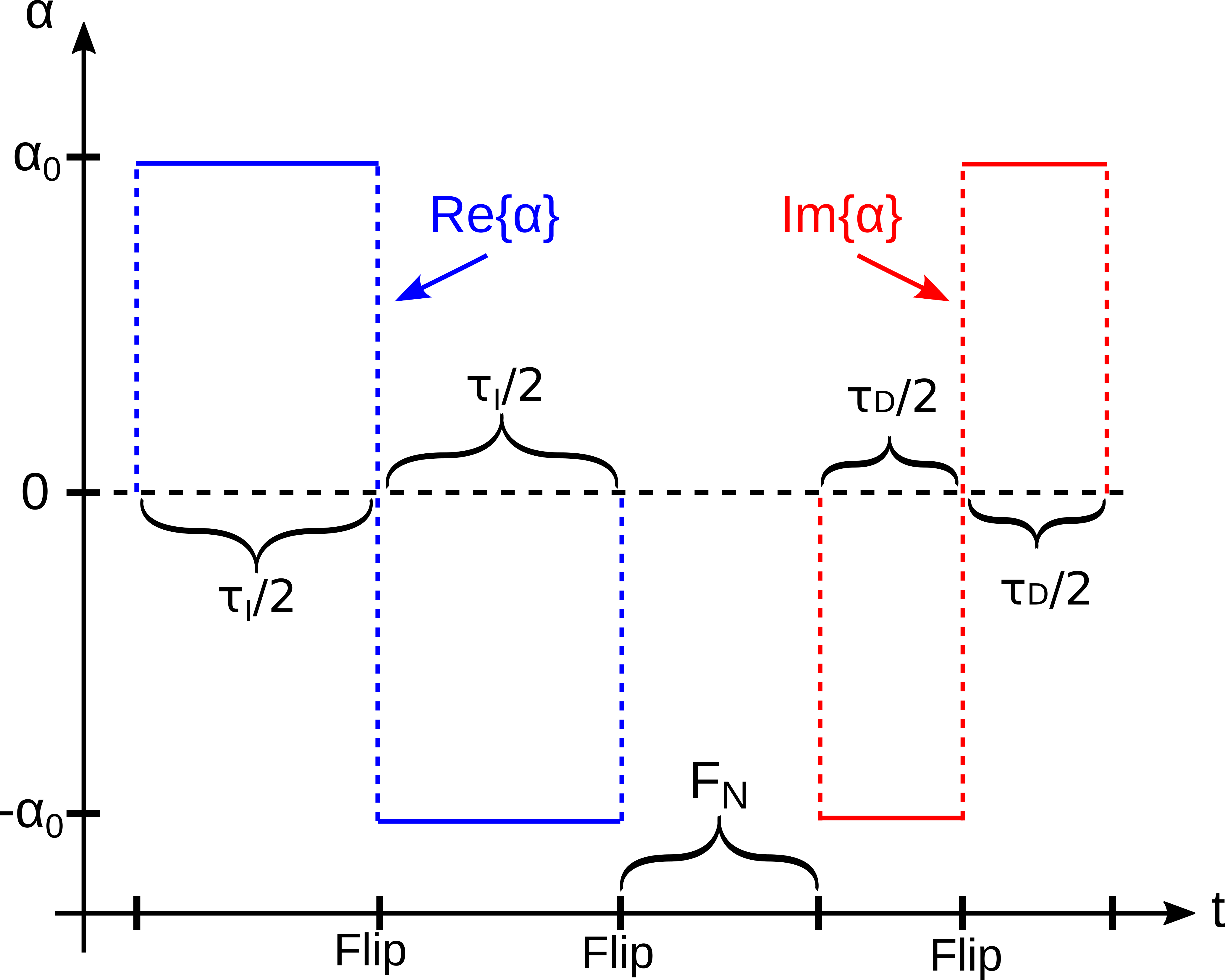}
		\caption{\textit{Ideal time dependence of $\alpha(t)$ that implements our protocol, with three qubit flips. The resulting state for $P_q=2\sqrt{\pi}$ can be seen in Fig. \ref{modsmeas:fig:alphaSeqWigner}. On the time axis we show the operations applied only to the qubits, where $F_N$ is the QFT.}}
		\label{modsmeas:fig:alphaSeq}
\end{figure*}

The idealized time-dependent displacement $\alpha(t)$ that implements our protocol is sketched in Fig. \ref{modsmeas:fig:alphaSeq}. The figure also shows the timing for qubit flips and the application of the QFT. Note that in an experimental implementation the idealized sharp steps of $\alpha(t)$ will be replaced by a smoothly varying function resulting from a realistic drive $\varepsilon(t)$ \citep{Eickbusch:2022}. The interaction time between the qubits and the oscillator is divided into $\tau_I$ and $\tau_D$, corresponding to the implementation of $U_I$ and $U_D$ respectively. These intervals are given by,
\begin{align}
	\tau_I &= \frac{2\sqrt{2}\pi}{\alpha_0 \chi P_q} \nonumber \\
	\tau_D &= \frac{\sqrt{2}P_q}{\alpha_0 \chi M },
\end{align}
where $\alpha_0 = \mathrm{max}\left(|\alpha(t)|\right)$ is the maximum displacement amplitude.
The peak spacing $P_q$ is set to $2\sqrt{\pi}$.

\begin{figure*}
	\centering		
	\includegraphics[width=0.7\textwidth]{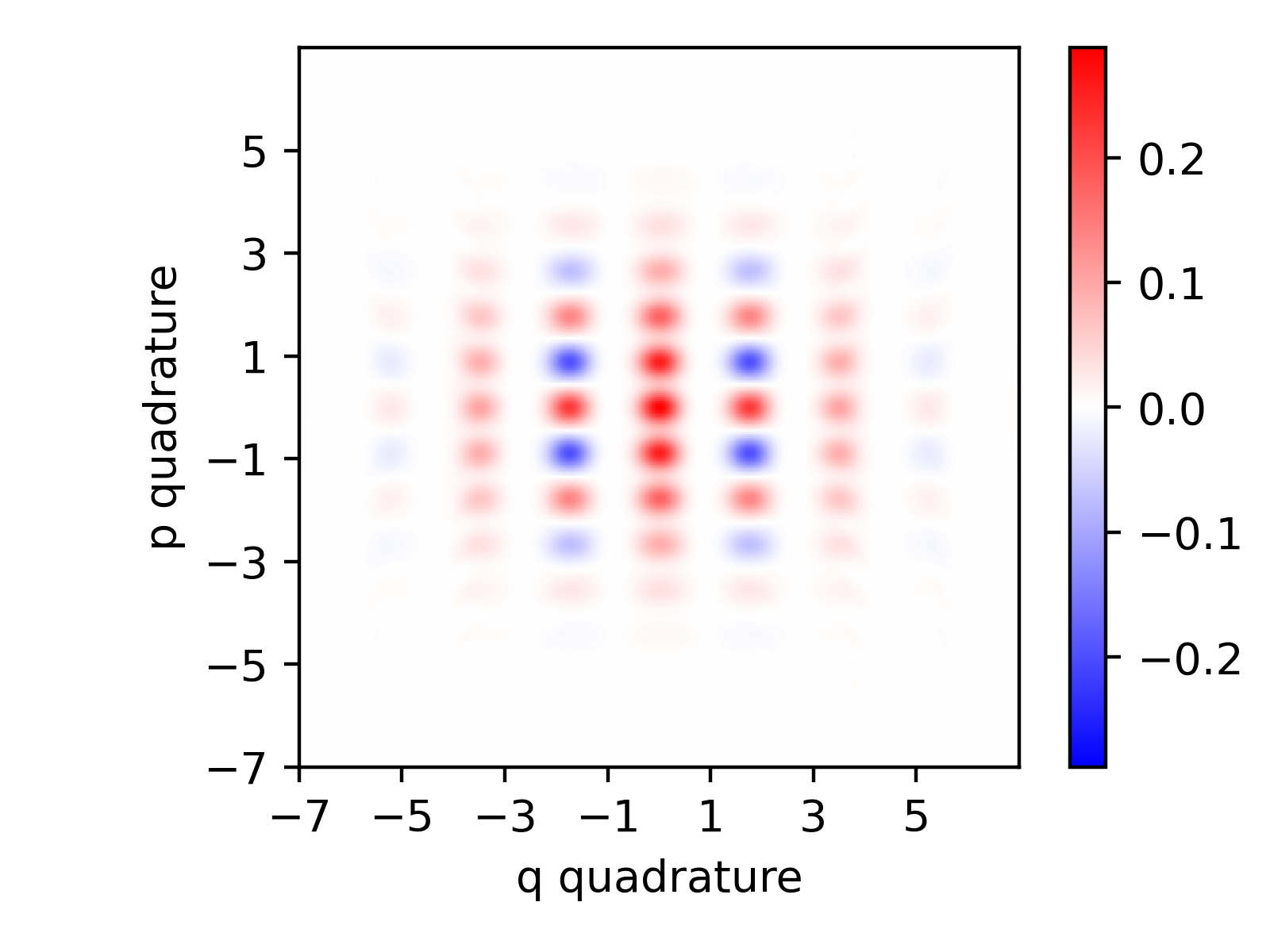}
		\caption{\textit{Wigner function of the oscillator state resulting from the sequence given in Fig. \ref{modsmeas:fig:alphaSeq} with 3 qubits. We have set $\chi=1$, $P_q=2\sqrt{\pi}$, and $\alpha_0=30$ (as in \cite{Eickbusch:2022}). The qubits are flipped 7 times. The slight slant is due to the interaction terms $a^\dagger a \sigma_z^{(n)}$ and vanishes if the qubits are flipped for more often, or $\alpha_0$ is increased.}}
		\label{modsmeas:fig:alphaSeqWigner}
\end{figure*}

From now on we set $\chi$ to 1, since in the absence of decoherence, $\chi$ is the only rate in our equation of motion. The noise rates, given in Table \ref{modmeas:table:noise}, are initially set to zero, but will later be defined relative to $\chi_\mathrm{max} = 2^{N-2} \chi$. 
In Fig. \ref{modsmeas:fig:alphaSeqWigner} we plot the Wigner function of the state resulting from the drive $\alpha(t)$ (Fig. \ref{modsmeas:fig:alphaSeq}) with 3 qubits and an initial squeezing of 10 dB. Instead of doing 3 qubit flips as in Fig. \ref{modsmeas:fig:alphaSeq}, the qubits are instead flipped 7 times. This reduces a distortion of the state resulting from the terms $a^\dagger a \sigma_z^{(n)}$. The produced state closely matches the target state shown in Fig. \ref{modsmeas:fig:wigner} a, with a slight slant that diminishes if the qubits are flipped more often during the protocol, or if $\alpha_0$ is increased.

\begin{figure*}
	\centering		
	\includegraphics[width=0.9\textwidth]{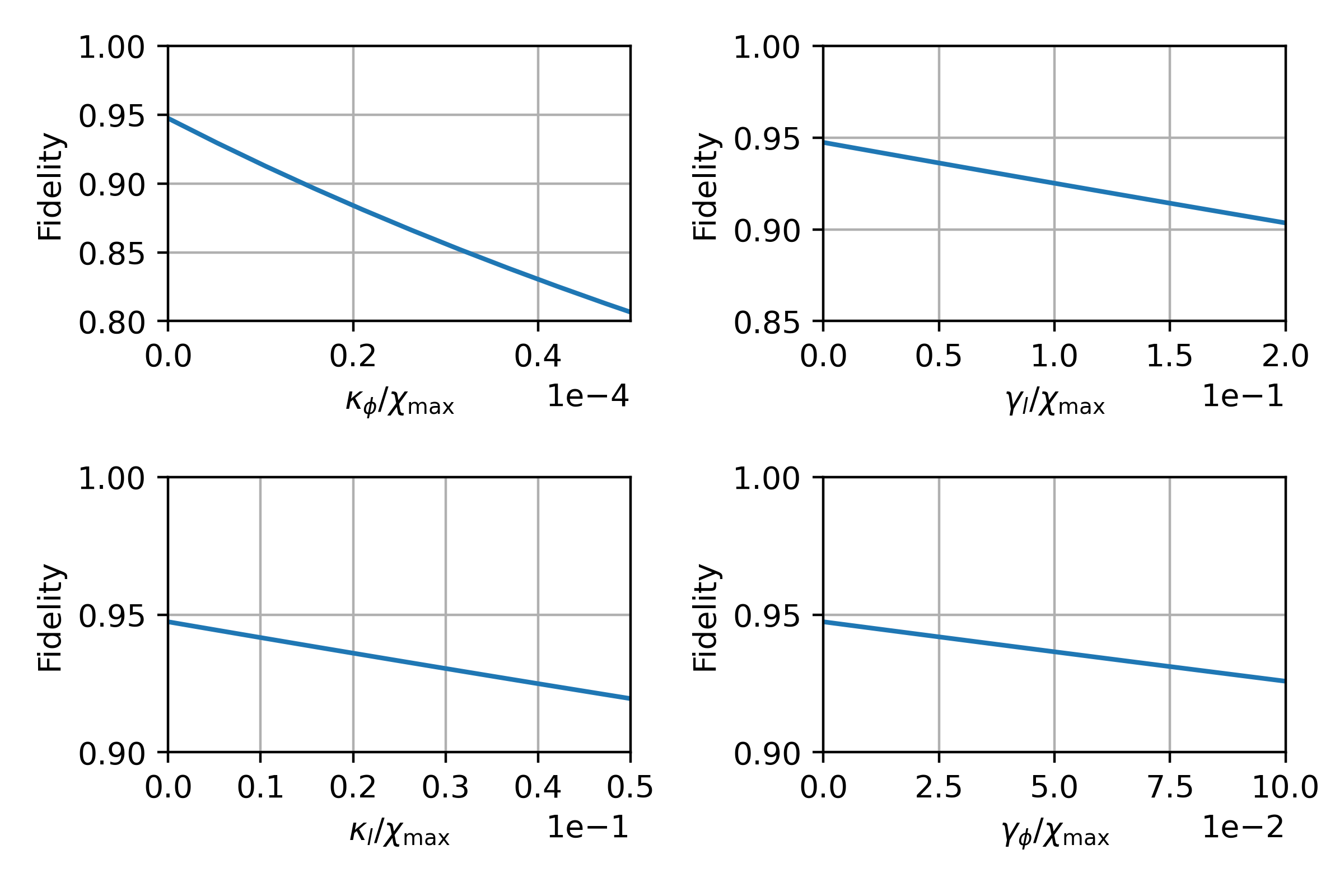}
		\caption{\textit{Fidelity between $G_0$ and $\rho_\kappa$ for various noise rates (see main text). On the x-axis we give the ratio between the noise rate and $\chi_{\mathrm{max}}$. $\chi$ is set to 1 and defines the time-scale of the simulation. $\alpha_0$ is set to 30. The plots do not include decoherence effects associated with the implementation of the QFT.}}
		\label{modsmeas:fig:fidelities}
\end{figure*}

We evaluate the robustness of the protocol to dephasing and loss during the controlled displacement sequence. Robustness is measured in terms of the fidelity between the produced state and the target GKP state. Assuming the protocol produces the state $\rho_0$ in the absence of noise, we determine the approximate GKP state $G_0$ (as given in Eq. \ref{intro:eq:approxGKP}) most similar to $\rho_0$, as measured by the fidelity. The state produced in the presence of noise with rate $\kappa$, is denoted $\rho_\kappa$. To gauge the sensitivity of the scheme to the noise sources listed in Table \ref{modmeas:table:noise}, we compute the fidelity between $\rho_\kappa$ and $G_0$ as we vary the noise rates individually. Plots of the fidelity against the noise rates are given in Fig. \ref{modsmeas:fig:fidelities}. The noise rates are chosen so as to be comparable with the rates given in \cite{Eickbusch:2022}. We set $\alpha_0=30$ (as in \cite{Eickbusch:2022}). The calculations do not include decoherence effects associated with the initialization of the qubits and oscillator (see Fig. \ref{modsmeas:fig:Prtcl2Circuit}), and the implementation of the QFT. We expect that the error channels associated with these operations might make our protocol more error prone than the comparable protocols \cite{Terhal:2016} and \cite{Eickbusch:2022}.

The protocol appears stable against oscillator loss and qubit noise. The controlled displacement is however sensitive to oscillator phase noise, due to the large displacement $\alpha_0$. The non-unit fidelity at zero noise results from the peaks not being perfect gaussians, finite squeezing of the initial wavefunction, and from a distortion generated by the $a^\dagger a \sigma_z^{(n)}$ interaction. 

\section{Conclusions}
We have presented a protocol for mapping a qudit state onto an oscillator using the periodic interpolating function $v(y)$ defined in Eq. \ref{modmeas:equation:interpolation}. The qudit was realized through a set of qubits, and we showed how the proposed mapping can be realized using controlled displacements. A circuit $U_v$ was constructed for the preparation of a qubit state suitable to prepare oscillator GKP states. Our analysis included computing the Wigner function of the oscillator state produced by the protocol with 3 and 4 qubits. 
The results showed a significant improvement in the sharpness of the approximate GKP state with 4 qubits compared to the 3-qubit implementation. This highlights the scalability and effectiveness of our protocol in producing high-quality GKP states with an increasing number of qubits.
Finally we investigated the feasibility of the protocol with 3 qubits, considering various noise sources such as qubit decay, qubit dephasing, oscillator loss and oscillator dephasing. 
Overall, our protocol offers a promising approach for efficiently generating highly squeezed GKP states, with robustness against most common noise sources, making it a viable tool for practical quantum computing.

\section{Acknowledgments}
This research is part of the Munich Quantum Valley, which is supported by the Bavarian state government with funds from the Hightech Agenda Bayern Plus. This work is also supported by the Danish National Research Foundation, Center for Macroscopic Quantum States (bigQ, DNRF142). This work is also supported by the European Quantum Flagship Quantum Secure Networks Partnership (EU QSNP) project.

\clearpage

\section{Appendix}

\subsection{Interpolating function}
\label{appendix:section:interpolater}

We evaluate the action of the displacement operator in Eq. \ref{protocol:eq:displacement}. We proceed by expanding the $Y_N$ operator in terms of its eigenvectors,
\begin{widetext}
\begin{align}
	D_{x}(s)\sum_{l\in K}v_{l}|x_{l}\rangle &= \sum_{l\in K}v_{l}e^{-i\frac{2\pi}{M}Y_{N}s}|x_{l}\rangle = \sum_{l\in K}v_{l}\sum_{n\in K}e^{-i\frac{2\pi}{M}ns}|y_{n}\rangle\langle y_{n}|x_{l}\rangle \nonumber \\
	 &= \frac{1}{\sqrt{M}}e^{-i\frac{\pi}{M}s}\sum_{l\in K}v_{l}\sum_{n\in K}e^{i2\pi\left(n-1/2\right)\left(-s-l+1/2\right)/M}|y_{n}\rangle \nonumber \\
	 &= e^{-i\frac{\pi}{M}s}\sum_{m\in K}\left[\frac{1}{M}\sum_{n,l\in K}v_{l}e^{i2\pi\left(n-1/2\right)\left(m-s-l\right)/M}\right]|x_{m}\rangle \nonumber \\
	 &= e^{-i\frac{\pi}{M}s}\sum_{m\in K}v(m-s)|x_{m}\rangle ,
\end{align}
where
\begin{align}
	v(y) = \frac{1}{M}\sum_{n,l\in K}v_{l}e^{i2\pi\left(n-1/2\right)\left(y-l\right)/M} .
\end{align}
\end{widetext}
By summing the geometric series one can readily verify that $v(k)$ equals $v_k$ for $k \in K$. Furthermore, calculating the Fourier coefficient of $v(y)$ corresponding to the frequency $m/M$ (with $m \in \mathbb{Z}$) over the range $R_0$, we find
\begin{align}
	&c_{m} = \frac{1}{\sqrt{M}}\int_{-M/2}^{M/2}dyv(y)e^{-i2\pi\frac{m}{M}y} \nonumber \\
&= \frac{1}{M^{3/2}}\sum_{n,l\in K}v_{l}e^{-i2\pi\left(n-1/2\right)l/M}\int_{-M/2}^{M/2}dye^{i2\pi y\left[n-1/2-m\right]/M} \nonumber \\
&= \frac{1}{\sqrt{M}}\sum_{l\in K}\sum_{n=-M/2}^{M/2-1}v_{l}e^{-i2\pi nl/M}\delta\left(n-m\right) \nonumber \\
&= \begin{cases}
\frac{1}{\sqrt{M}}\sum_{l\in K}v_{l}e^{-i2\pi ml/M} & \mathrm{if} \hspace{0.2cm} m \in [-M/2;M/2-1] ,\\
0 & \mathrm{otherwise} ,
\end{cases}
\end{align}
where the $\delta$ is a Kronecker $\delta$. So we can write the Fourier coefficient of $v(y)$ associated with frequency $m/M$ as,
\begin{align}
	c_{m} &=
	\begin{cases}
        \langle \phi_m | v\rangle  & \mathrm{if} \hspace{0.2cm} m \in [-M/2;M/2-1] , \\
        0 & \mathrm{otherwise} ,
    \end{cases}
\end{align}
where we defined the plane wave,
\begin{align}
	|\phi_m\rangle = \frac{1}{\sqrt{M}}\sum_{l\in K}e^{i2\pi ml/M} |x_l\rangle .
\end{align}
It follows, that as long as $\langle \phi_m|v \rangle$ is only large for $|m| \ll M/2$, then the frequencies of $v(y)$ will be significantly below 1/2. The interpolation $v(y)$ of the points $(k,v_k)$ can then be considered reasonable, in the sense that there will be little oscillatory behaviour between the interpolated points $(k,v_k)$. \\

\subsection{Frame transformations}
\label{appendix:section:frameTrans}

We assume that the density matrix $\rho$ evolves according to a master
equation in Lindblad form,
\begin{align}
\partial_{t}\rho=-i[H/\hbar,\rho]+\sum_{i}\kappa_{i}D[L_{i}]\rho
\end{align}
with $D[L_{i}]\rho=L_{i}\rho L_{i}^{\dagger}-\frac{1}{2}\left(L_{i}^{\dagger}L_{i}\rho+\rho L_{i}^{\dagger}L_{i}\right)$
and $L_{i}$ is the jump operator and $\kappa_{i}$ is the jump rate.
We make the frame transformation,
\begin{align}
\tilde{\rho}=U\rho U^{\dagger}
\end{align}
thereby obtaining the new equation of motion,
\begin{align}
\partial_{t}\tilde{\rho}=-i[i\dot{U}U^{\dagger}+U\left(H/\hbar\right)U^{\dagger},\tilde{\rho}]+\sum_{i}\kappa_{i}D[UL_{i}U^{\dagger}]\tilde{\rho} .
\end{align}

\subsection{Hamiltonian}
We use the dispersive Hamiltonian with a drive $\varepsilon(t)$,
\begin{align}
H/\hbar &=\sum_{n=1}^{N}\left\{ \chi^{(n)}a^{\dagger}a\frac{\sigma_{z}^{(n)}}{2}+\frac{1}{2}\omega_{q}^{(n)}\sigma_{z}^{(n)}\right\} \nonumber \\ &+\varepsilon^{*}(t)a+\varepsilon(t)a^{\dagger}+\omega_{o}a^{\dagger}a ,
\end{align}
where $\sigma_{z}=|0\rangle\langle0|-|1\rangle\langle1|$ and $\omega_o$ is the angular frequency of the oscillator. We change
to the rotating frame,
\begin{align}
U_{1}=e^{i\omega_{o}a^{\dagger}at+i\frac{1}{2}\sum_{n=1}^{N}\omega_{q}^{(n)}\sigma_{z}^{(n)}t} ,
\end{align}
thereby obtaining the Hamiltonian,
\begin{align}
&H_{1}/\hbar = i\dot{U_{1}}U_{1}^{\dagger}+U_{1}\left(H/\hbar\right)U_{1}^{\dagger} \nonumber \\
&=\sum_{n=1}^{N}\frac{1}{2}\chi^{(n)}a^{\dagger}a\sigma_{z}^{(n)}+\varepsilon^{*}(t)e^{-i\omega_{o}t}a+\varepsilon(t)e^{i\omega_{o}t}a^{\dagger} .
\end{align}
We change to a displaced frame,
\begin{align}
U_{2}=e^{\alpha^{*}(t)a-\alpha(t)a^{\dagger}} .
\end{align}
The new Hamiltonian is,
\begin{align}
H_{2}/\hbar&=i\dot{U_{2}}U_{2}^{\dagger}+U_{2}\left(H_{1}/\hbar\right)U_{2}^{\dagger} \nonumber \\
&=\sum_{n=1}^{N}\bigg\{ \frac{1}{2}\chi^{(n)}a^{\dagger}a\sigma_{z}^{(n)}+\frac{1}{2}\chi^{(n)}|\alpha(t)|^{2}\sigma_{z}^{(n)} \nonumber \\
&+\frac{1}{2}\chi^{(n)}\left(\alpha(t)a^{\dagger}+\alpha^{*}(t)a\right)\sigma_{z}^{(n)}\bigg\} \nonumber \\
&+\phi(t)^{*}a+\phi(t)a^{\dagger}+\beta(t) , 
\end{align}
where
\begin{align}
\beta(t)&=\varepsilon^{*}(t)e^{-i\omega_{o}t}\alpha(t)+\varepsilon(t)e^{i\omega_{o}t}\alpha^{*}(t) , \\
\phi(t)&=\varepsilon(t)e^{i\omega_{o}t}-i\dot{\alpha}(t) .
\end{align}

\subsection{Lindbladians}

The corresponding transformations of the Lindbladians, resulting from the frame
transformations, are,

\vspace{0.3cm}
\paragraph*{Oscillator loss}
\begin{align}
D[U_{2}U_{1}aU_{1}^{\dagger}U_{2}^{\dagger}]\rho=D[a]\rho-\frac{1}{2}\alpha(t)\left[a^{\dagger},\rho\right]+\frac{1}{2}\alpha(t)^{*}\left[a,\rho\right] .
\end{align}

\vspace{0.3cm}
\paragraph*{Oscillator dephasing}

\begin{align}
D[U_{2}U_{1}a^{\dagger}aU_{1}^{\dagger}U_{2}^{\dagger}]\rho=D\left[(a^{\dagger}+\alpha(t)^{*})(a+\alpha(t))\right]\rho .
\end{align}

\vspace{0.3cm}
\paragraph*{Qubit decay}

\begin{align}
D[U_{2}U_{1}\sigma_{-}^{(n)}U_{1}^{\dagger}U_{2}^{\dagger}]\rho=D[\sigma_{-}^{(n)}]\rho .
\end{align}
where $\sigma_{-}=|1\rangle\langle0|$. 


\vspace{0.3cm}
\paragraph*{Qubit dephasing}

\begin{align}
D[U_{2}U_{1}\frac{1}{2}\sigma_{z}^{(n)}U_{1}^{\dagger}U_{2}^{\dagger}]\rho=D[\frac{1}{2}\sigma_{z}^{(n)}]\rho
\end{align}

\subsection{Master equation in frame $\mathcal{D}$}
\label{appendix:section:masterEq}

Letting $\rho_{\mathcal{D}}=U_{2}U_{1}\rho U_{1}^{\dagger}U_{2}^{\dagger}$,
then the master equation in this frame states,
\begin{align}
&\partial_{t}\dot{\rho_{\mathcal{D}}}=-i\left[H_{2}/\hbar+\frac{1}{2}i\kappa_{l}\left(\alpha(t)^{*}a-\alpha(t)a^{\dagger}\right),\rho_{\mathcal{D}}\right] \nonumber \\
&+\kappa_{l}D[a]\rho_{\mathcal{D}}+2\kappa_{\phi}D\left[(a^{\dagger}+\alpha(t)^{*})(a+\alpha(t))\right]\rho_{\mathcal{D}} \nonumber \\
&+\sum_{n=1}^{N}\left\{ \gamma_{l}D[\sigma_{-}^{(n)}]\rho_{\mathcal{D}}+2\gamma_{\phi}D[\frac{1}{2}\sigma_{z}^{(n)}]\rho_{\mathcal{D}}\right\} ,
\end{align}
from which we obtain the effective Hamiltonian,
\begin{align}
&H_{\mathrm{eff}}/\hbar=H_{2}/\hbar+\frac{1}{2}i\kappa_{l}\left(\alpha(t)^{*}a-\alpha(t)a^{\dagger}\right) \nonumber \\
&=\sum_{n=1}^{N}\bigg\{ \frac{1}{2} \chi^{(n)}a^{\dagger}a\sigma_{z}^{(n)} + \frac{1}{2}\chi^{(n)}|\alpha(t)|^{2}\sigma_{z}^{(n)} \nonumber \\ 
&+\frac{1}{2} \chi^{(n)}\left(\alpha(t)a^{\dagger}+\alpha^{*}(t)a\right)\sigma_{z}^{(n)}  \bigg\} \nonumber \\
&+\left(\phi(t)^{*}+\frac{1}{2}i\kappa_{l}\alpha(t)^{*}\right)a+\left(\phi(t)-\frac{1}{2}i\kappa_{l}\alpha(t)\right)a^{\dagger}.
\end{align}
We then require that the drive $\varepsilon(t)$ (approximately) satisfy
the following equation,
\begin{align}
\phi(t)-\frac{1}{2}i\kappa_{l}\alpha(t)=\varepsilon(t)e^{i\omega_{o}t}-i\dot{\alpha}(t)-\frac{1}{2}i\kappa_{l}\alpha(t)=0.
\end{align}
The effective Hamiltonian is then,
\begin{align}
&H_{\mathrm{eff}}/\hbar=\sum_{n=1}^{N}\bigg\{ \frac{1}{2}\chi^{(n)}a^{\dagger}a\sigma_{z}^{(n)} + \frac{1}{2}\chi^{(n)}|\alpha(t)|^{2}\sigma_{z}^{(n)} \nonumber \\
&+\frac{1}{2}\chi^{(n)}\left(\alpha(t)a^{\dagger}+\alpha^{*}(t)a\right)\sigma_{z}^{(n)} \bigg\} .
\label{appendix:eq:effH}
\end{align}

\clearpage
\bibliography{bibliography.bib}

\end{document}